# Analytical Solution Describing the Periodicity of the Elements in the Periodic System

Jozsef Garai
Department of Earth Sciences, Florida International University, University Park PC 344, Miami, FL 33199
E-mail: jozsef.garai@fiu.edu

Mathematical formula describing the periodicity of the elements in the periodic system is presented.

---

In the periodic system (Fig. 1) the number of electrons in the completely filled outer shell (full-shell), and the atomic numbers of the elements with full-shell follow the sequences

$$2, 8, 8, 18, 18, 32, 32\ldots, \qquad (1)$$

and

$$2, 10, 18, 36, 54, 86, 118\ldots \qquad (2)$$

respectively. In the structural development of the nucleus the same sequences should be present regarding to the number of protons. No numerical solutions describing these periodicities are known[3].

Quantum mechanics does not give a complete explanation for the observed periodicity of the elements[4]. The completed shells predicted by quantum mechanics are consistent only with the first two periods. Later periods do not correlate with the quantum mechanical predictions. The repeated pattern of the periods is not only marked by the same number of electrons/protons in the fully developed shell but also by the repeated pattern of the physical characteristics of the elements (Fig. 2).

Double tetrahedron shape with alternately arranged protons and neutrons in face-centered cubic lattice has been proposed for the structure of the nucleus[6]. This structure is able to reproduce all of the quantum numbers without discrepancy. The generated numbers from the structure not only numerically match but also bear the same physical meaning as their corresponding quantum counterpart. Additionally the sequence of the structural development of the double tetrahedron is consistent with the periodicity of the periodic system. This consistency allows one to calculate the number of protons in the outer shell and the total number of protons in the nucleus and by that describe the periodicity of the elements. The analytical solution is derived here.

The general description of the repeated periods in the periodic system can be given by the sequence

$$1, 2, 2, 3, 3, 4, 4, 5, 5, \ldots \qquad (3)$$

I will call the numbers in Eq. (3) to sequence numbers of the periods. The relationship between the period and sequence numbers can be given as:

$$m = n + \frac{[(-1)^n - 1](n-1)}{4} - \frac{[(-1)^n + 1](n-2)}{4} \tag{4}$$

where n is the number of the period and m is the sequence number.

The number of protons in the $k^{th}$ layer of a tetrahedron $[N^{proton}_{\Delta-layer}(k)]$ can be calculated from the triangular number[7,8] $[Tr(k)]$ (Fig. 3. a)

$$N^{proton}_{\Delta-layer}(k) = \frac{Tr(k)}{2}, \tag{5}$$

where

$$Tr(k) = \frac{k}{2}(k+1). \tag{6}$$

In each structural step of the development of the tetrahedron, starting with a four unit 'core' tetrahedron, one additional layer is added to the two outer sides (Fig. 3 b-d). The number of nucleons $[N^{nucleon}_{\Delta-full-shell}]$ in these outer layers or shells is the sum of the two consecutive triangular numbers covering the two sides of the tetrahedron.

$$N^{nucleon}_{\Delta-full-shell} = Tr(k) + Tr(k-1) = \frac{k}{2}(k+1) + \frac{k-1}{2}k = k^2 \tag{7}$$

Each shell of the tetrahedron corresponds to two tetrahedron layer Eq. (7). The relationship between the tetrahedron layers and the sequence numbers is

$$k = 2m. \tag{8}$$

Substituting the sequence number into Eq. (7) gives the number of nucleons in a given shell as:

$$N^{nucleon}_{\Delta-full-shell}(m) = (2m)^2 = 4m^2 \tag{9}$$

The number of protons in the completely developed shell $[N(m)^{proton}_{\Delta-full-shell}]$ is then

$$N(m)^{proton}_{\Delta-full-shell} = \frac{N^{nucleon}_{\Delta-full-shell}(m)}{2} = 2m^2 \tag{10}$$

Substituting the sequence number from Eq. (4) gives the general formula for the number of protons in the completely developed outer shell of the nucleus for any period.

$$N(n)^{proton}_{full-shell} = 2\left\{n + \frac{[(-1)^n - 1](n-1)}{4} - \frac{[(-1)^n + 1](n-2)}{4}\right\}^2 \tag{11}$$

Formula giving the total number of protons in the nucleus with completely developed shells can be derived in a similar manner. The total number of nucleons in a tetrahedron with k layers can be determined by its tetrahedral number[7,9] $[Th(k)]$

$$Th(k) = \frac{k}{6}(k+1)(k+2) \tag{12}$$

Substituting the sequence number from Eq. (8) gives the number of nucleons in a tetrahedron for sequence (m) as:

$$\text{Th}(m) = \frac{m}{3}(2m+1)(2m+2) \tag{13}$$

and the number of protons as

$$N_{\Delta-\text{total}}^{\text{proton}}(m) = \frac{\text{Th}(m)}{2} = \frac{m}{6}(2m+1)(2m+2) = \frac{2m^3}{3} + m^2 + \frac{m}{3} \tag{14}$$

Substituting the sequence number of the periods from Eqs. (4) into (14) gives the total number of protons in one tetrahedron for any periods

$$N_{\Delta-\text{total}}^{\text{proton}}(n) = \frac{2}{3}\left\{n + \frac{\left[(-1)^n - 1\right](n-1)}{4} - \frac{\left[(-1)^n + 1\right](n-2)}{4}\right\}^3 +$$

$$+ \left\{n + \frac{\left[(-1)^n - 1\right](n-1)}{4} - \frac{\left[(-1)^n + 1\right](n-2)}{4}\right\}^2 + \tag{15}$$

$$+ \frac{1}{3}\left\{n + \frac{\left[(-1)^n - 1\right](n-1)}{4} - \frac{\left[(-1)^n + 1\right](n-2)}{4}\right\}$$

The total number of protons in the double tetrahedron nucleus with completely developed shells is the atomic number of the elements with full-shell $\left[Z(n)^{\text{full-shell}}\right]$. Based on the alternate development of the double tetrahedron (Fig. 4) the atomic number of an element with full shell can be calculated for any period as

$$Z(n)^{\text{full-shell}} = 2N_{\Delta-\text{total}}^{\text{proton}}(n) - N(n^{\text{even}})_{\text{full-shell}}^{\text{proton}} - 2 \tag{16}$$

The formula

$$\frac{(-1)^n + 1}{2} \tag{17}$$

can be used to generate 0 for add periods and 1 for even number periods, and

$$N(n^{\text{even}})_{\text{full-shell}}^{\text{proton}} = \left[(-1)^n + 1\right] N(n)_{\text{full-shell}}^{\text{proton}} = \left[(-1)^n + 1\right]\left\{n - \frac{\left[(-1)^n + 1\right](n-2)}{4}\right\}^2. \tag{18}$$

Substituting Eqs. (15), and (18) into Eq. (16) gives the total number of protons with completely developed shells for any periods

$$Z(n)^{full-shell} = \frac{4}{3}\left\{n + \frac{[(-1)^n - 1](n-1)}{4} - \frac{[(-1)^n + 1](n-2)}{4}\right\}^3 +$$

$$+ 2\left\{n + \frac{[(-1)^n - 1](n-1)}{4} - \frac{[(-1)^n + 1](n-2)}{4}\right\}^2 +$$

$$+ \frac{1}{3}\left\{2n + \frac{[(-1)^n - 1](n-1)}{2} - \frac{[(-1)^n + 1](n-2)}{2}\right\} - 2 - \qquad (19)$$

$$- [(-1)^n + 1]\left\{n - \frac{[(-1)^n + 1](n-2)}{4}\right\}$$

The derived analytical solution for the number of protons in a completely developed outer shell Eq. (11), and for the total number of protons in the double tetrahedron nucleus Eq. (19) reproduces the periodicity of the elements in the periodic system Eqs. (1), and (2) respectively.

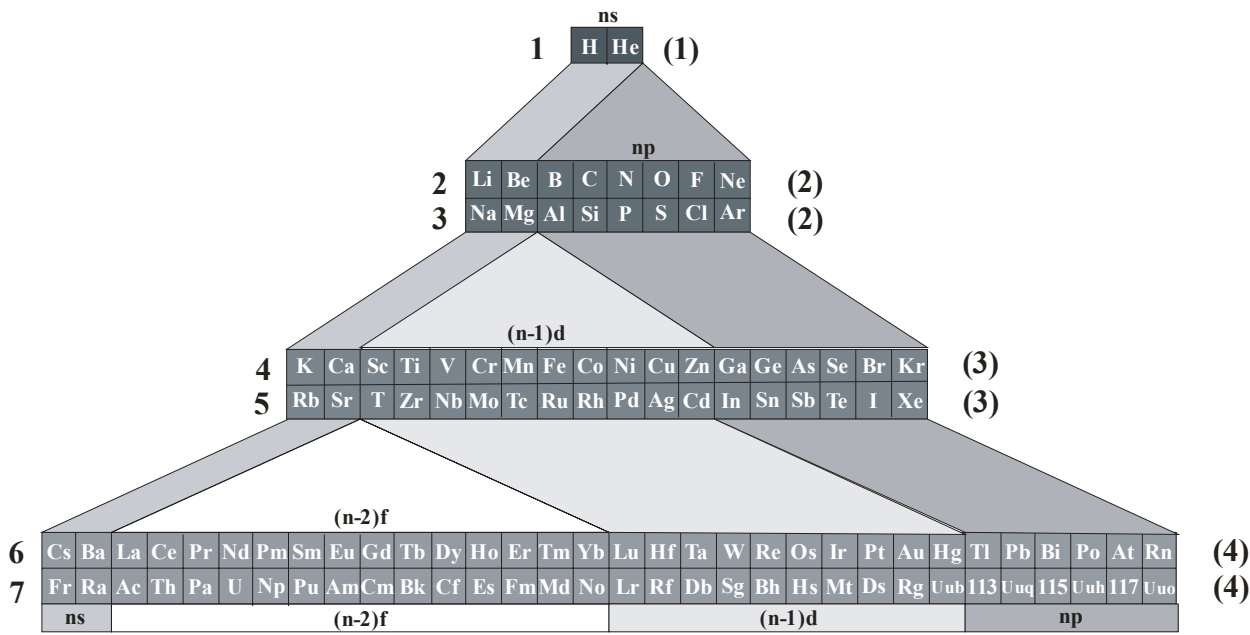

**Figure 1.** Pyramid arrangement of the elements in the periodic system[1] as developed by William B Jensen. The periods are numbered at the left while the sequence numbers (explained in the text.) are given at the right in parentheses. Elements 112, 114, 116, and 118 has been reported, but not fully authenticated and the assigned names are only provisional[2]. Elements 113, 115, and 117 have yet to be discovered.

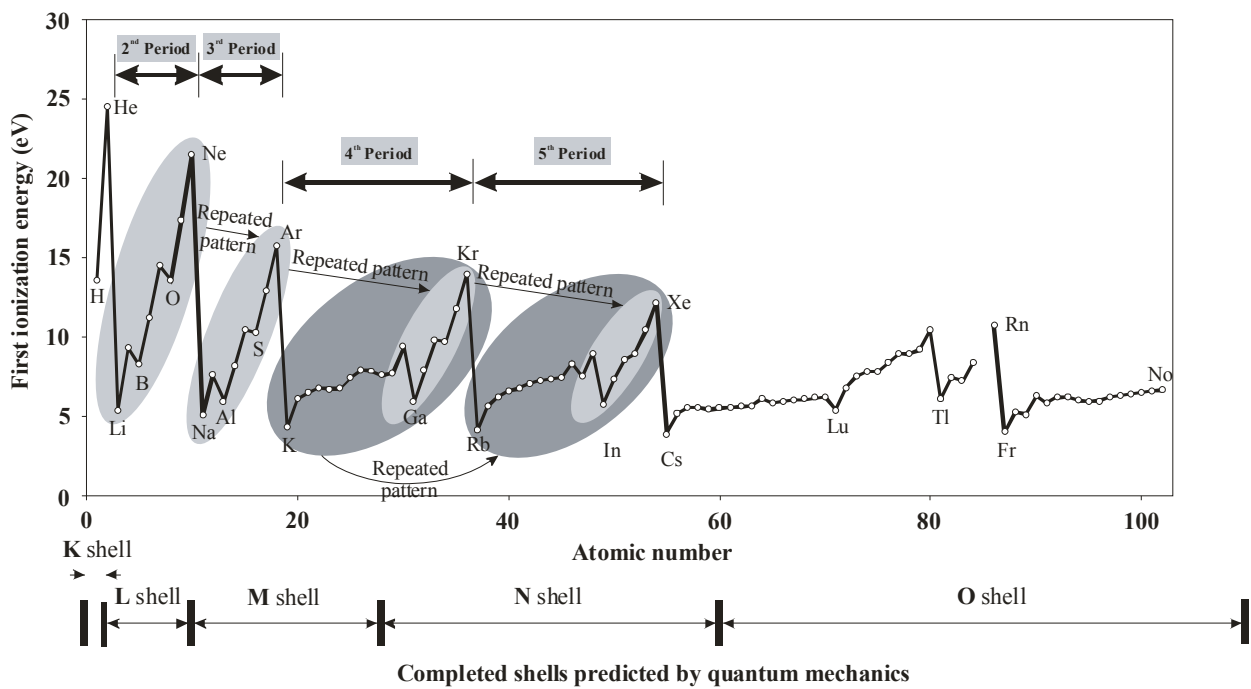

**Figure 2.** The first ionization energy of the elements[5] is given as an example to illustrate the double periodic pattern in the physical characteristics of the elements.

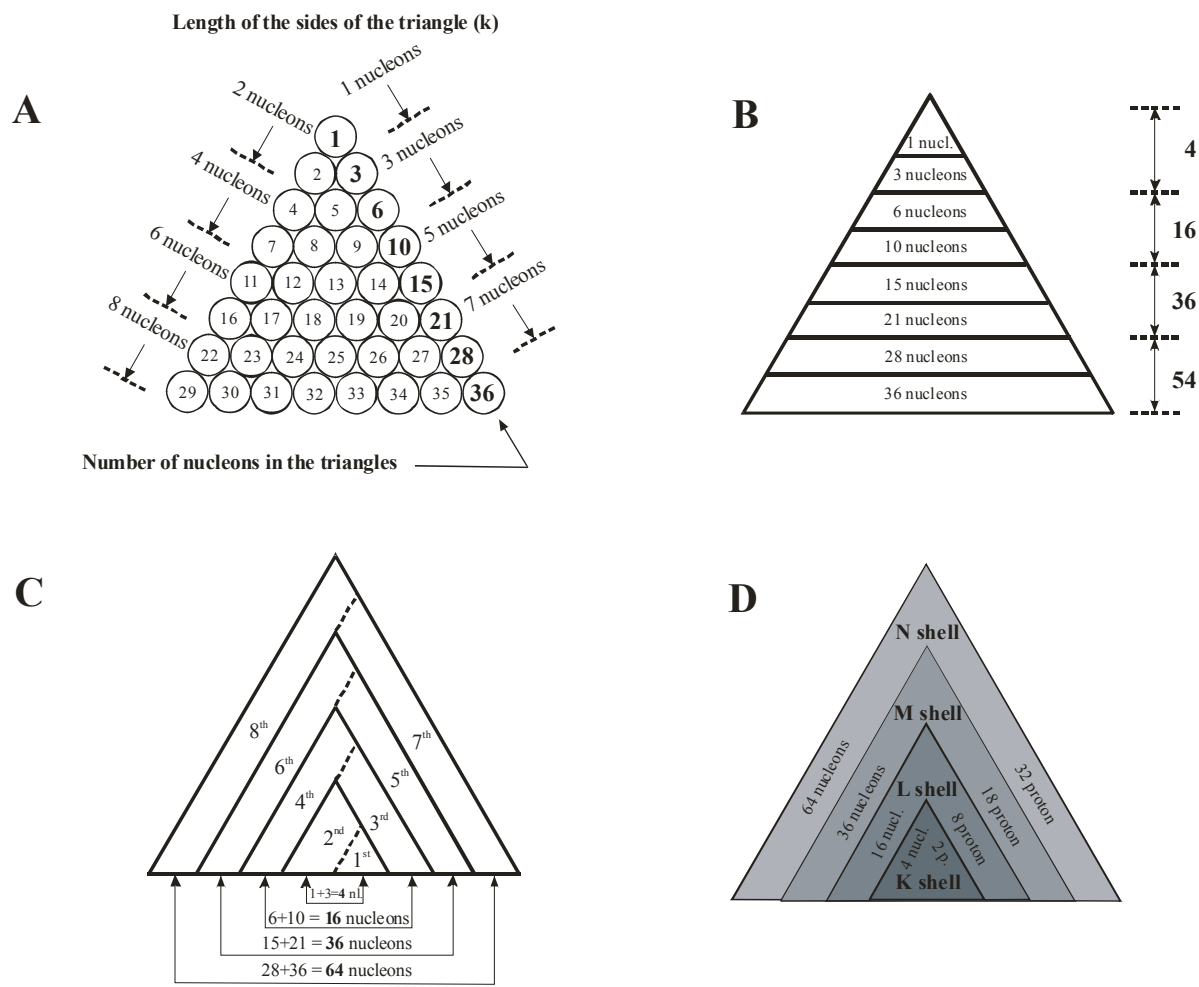

**Figure 3.** Number of nucleons and protons in a tetrahedron is identical with the quantum mechanical predictions. (a) The number of nucleons in the triangles. (b) Forming a tetrahedron by the staking the triangular layers. The nucleons are arranged in face center cubic lattice. (c) Rearranging the triangular layers in order to form shells. (d) The number of nucleons and protons in the shells.

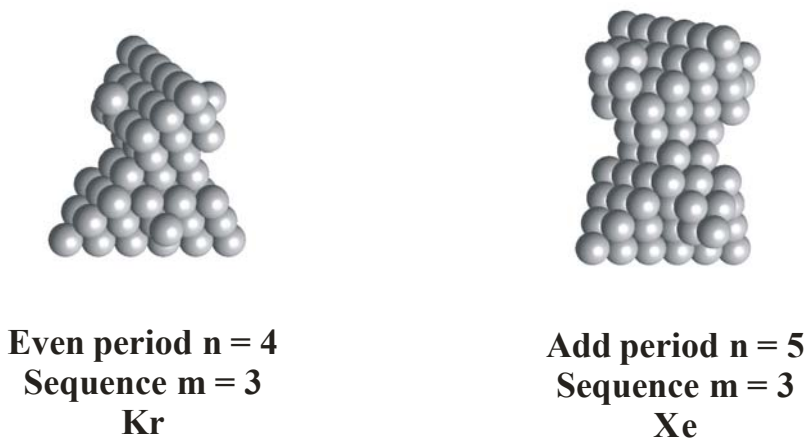

**Even period n = 4**
**Sequence m = 3**
**Kr**

**Add period n = 5**
**Sequence m = 3**
**Xe**

**Figure 4.** Examples of completely developed double tetrahedron nucleuses are shown for even and add periods. The 3D images of Kr and Xe.